\documentclass[12pt,a4paper]{article}

\usepackage{amsmath}
\usepackage{amsthm}
\usepackage{amscd}
\usepackage{amsxtra}
\usepackage{upref}
\usepackage{amsfonts}
\usepackage{amssymb}
\usepackage{graphicx}
\usepackage{url}
\usepackage{cite}
\usepackage{pifont}

\def\beq{\begin{equation}}
\def\eeq{\end{equation}}
\def\bea{\begin{eqnarray}}
\def\eea{\end{eqnarray}}

\theoremstyle{definition}

\numberwithin{equation}{section}

\title{Birth of the Universe from the Multiverse}
\author{Laura Mersini-Houghton\\
Department of Physics and Astronomy, UNC Chapel Hill, NC 27599, USA.}
\begin{document}
\maketitle
\date{\today}
\begin{abstract}
This proceeding is based on a talk I gave at the 13th Marcel Grossmann Meeting in Stockholm Sweden. 
\end{abstract}
\section{}
Compactification of extra dimensions in string theory leads to a vast number of $(3+1)$ dimensional worlds, (about $10^{500}$ so far), coined the landscape. At the time of the discovery of the landscape, the question which one of these worlds is our universe seemed hopeless. Many argued that the vastness of the landscape undermines the very foundations of string theory for two reasons: (i) the theory seemed unfalsifiable since for every observation we could find a matching world on the landscape; (ii) the method advocated at the time for making sense of this landscape was the anthropic principle. The former objection implied string theory can not be scientific. The latter concern is that anthropics do not help scientific inquiry and rigor ~\cite{laurafred} but rather it may seem to push some version of creationism to the next level. For these reasons the whole field of string theory and also, of cosmology that relied on it for answers about fundamental questions such as the origins of the universe, seemed to be in deep crisis at the beginning of the millenia. 

In 2004 I proposed to view the landscape as the configuration space of the initial conditions  of the universe and to derive these initial conditions by the method of quantum cosmology. The idea of this proposal was that we allow the wavefunction of the universe to propagate on the landscape ~\cite{laura1}. Since then, I have been advocating that the question of the origins can only be addressed if we have an ensemble of possible initial conditions, a.k.a a multiverse ~\cite{lauramulti}. In this theory, the landscape is the working model of the ensemble of initial conditions. The idea of placing the wavefunction of the universe on the landscape of string theory, provided a formalism by which we could calculate and solve quantum equations- known as the Wheeler De Witt (WDW) equation- on the N-body landscape, thus deriving the most probable initial conditions for the origins of the universe in that structure. The WDW equation is:  $H \Psi[a,\phi] = 0$, where $H = H_g + H_{\phi}$ is the total hamiltonian of the system with $H_g, H_{\phi}$ the components that depends on the gravitational and scalar field degrees of freedom respectively. $\Psi[a, \phi]$ is the wavefunctional of the universe operating on a minisuperspace defined by the two variables: the metric of FRW $3-$geometries given by the scalar factor $a(t)$ with line element
\begin{equation}
ds^2= -N^2 dt^2 + a^2(t) d{\bf x}^2 ,
\label{eq:lineelt}
\end{equation} 
 and, the scalar field $\phi$ which is the collective moduli field that parametrizes the landscape~\cite{douglas} with energy $V(\phi)$. Due to the structure of the landscape where most of the internal degrees of freedom are Gaussian peaked around a mean value, the parametrization of the landscape by the collective coordinate $\phi$ and potential energy $V(\phi)$ is justified.
The WDW equation reads
\begin{eqnarray}
& &{\hat {\cal H}}\Psi(a,\phi) = 0 ~{\rm with} \nonumber \\
& &\hat{{\cal H}}=\frac{1}{2e^{3\alpha}}\left[\frac{4\pi}{3M_p^2}
\frac{\partial^2}{\partial\alpha^2}-
\frac{\partial^2}{\partial\phi^2}+e^{6\alpha}V(\phi)\right].
\label{eq:wdweq}
\end{eqnarray}
Here the scale factor $a$ has been written as $a=e^{\alpha}$  and all the information about the landscape is contained in $V(\phi)$. After solving the WDW equation on the landscape we found that the most probable universe is the one that sits at the lowest energies on the landscape ~\cite{laura1}. The reason for the selection of low energy vacua in the above first stage of derivation~\cite{laura1}, is that we needed to have included decoherence among the various solutions of the WDW equation in the formalism, in order to hope for deriving a realistic answer for a universe that starts as a quantum wavepacket but then grows and undergoes the quantum to classical transition.

In 2005 with R.Holman we included decoherence into the above proposal by taking into account the backreaction of the long wavelength quantum fluctuations $f_n$ of the field and of the metric comprising the bath, onto the wavefunction of the universe being the system~\cite{laurarich} $h_{ij} = a^2 \left(\Omega_{ij}+\epsilon_{ij}\right),\ \phi=\phi_0+\sum_n f_n(a) Q_n$. 
 Our initial WDW equation became a Master equation $H\Psi[a,\phi,f_n] = -\Sigma H_{n} \Psi[a,\phi,fn]$ operating on an infinite dimensional medi-superspace $(a, \phi, f_n)$ where $H_n$ contains the backreaction of fluctuation modes $f_n$ . We solved this equation as an N-body problem in an infinite dimesnioanl medi superspace \cite{laurarich}, without reducing it to a 2-body tunneling problem. When solved correctly, due to destructive interference from multiple scattering of the wavefunction on the landscape, the wavefunction undergoes Anderson localization ~\cite{anderson}.
  The full quantized Hamiltonian $\hat{H} = \hat{H}_0 + \sum_n {\hat{H}}_n$ then acts on the wavefunction which can be approximately written as $\Psi \sim \Psi_0 (a, \phi_0) \prod_n \psi_n (a, \phi, f_n)$ 
leading to the Master equation $\hat{H}_0 \Psi_0 (a, \phi_0) = \left(-\sum_n \langle  \hat{H}_n\rangle\right) \Psi_0 (a, \phi_0)$.
The angular brackets denote expectation values in the wavefunction $\psi_n$, $m^2 \simeq V''(\phi)$ is the curvature around the vacua where the wavefunction is localized, and $
\hat H_n = -\frac{\partial^2}{\partial f_n^2} + e^{6 \alpha} \left( m^2 +e^{-2 \alpha} \left(n^2-1\right)\right) f_n^2$
We solved the Master equation as an N-body problem and estimated the probability by the quantum mechanics expression $P \simeq |\Psi|^2$ . As an aside, this system is very similar and in the same universality class as quantum dots and spin glass in condensed matter. 

The result of this derivation, with decoherence included, was that the most probable universes are the ones that start at high energy vacua since only they can survive the backreaction of massive fluctuations. This theory thus lead to the emergence of a quantum dynamic selection criterion for the origins of our universe.
The effect of the 'bath' of long wavelength fluctuations onto the system-the wavefunction of the universe- is to trigger decoherence. Decoherence and entanglement are closely related. Calculating the effect of the latter onto our observable sky, led us to a series of observational predictions \cite{avatars}. The backreaction term incldued in the Master equation, arising from the 'bath' of fluctuations, induces an energy shift to the classical path of the wavepacket. This energy shift becomes a nonlocal energy correction to the Friedman equation of expansion and to the gravitational potential of our universe through a Poisson equation. With T. Takahashi, we calculated the nonlocal energy and graviational potential corrections due to superhorizon entanglement in 2006 ~\cite{avatars}, by methods of 2nd order WKB. The highly nonlocal correction we derived for the Friedman equation from the shift of the wave's classical path is required for consistency with Einstein equations and is given by $H^2 = \frac{1}{3 M_{\rm P}^2}
\left[V(\phi)+\frac{1}{2} \left(\frac{V(\phi)}{3 M_{\rm
P}^2}\right)^2 F(b,V)\right]\equiv \frac{V_{\rm eff}}{3 M_{\rm P}^2}$ with
\begin{eqnarray}
 \label{eq:corrfactor}
F(b,V) 
&=& 
\frac{3}{2} \left(
2+\frac{m^2M^2_{\rm P}}{V}\right)
\log \left( \frac{b^2 M_{\rm P}^2}{V}\right)\nonumber \\
&-&
\frac{1}{2} 
\left(1+\frac{m^2}{b^2}\right) \exp\left(-3\frac{b^2
M_{\rm P}^2}{V}\right). 
\end{eqnarray}
All surviving branches provide the current multiverse. The correction to our gravitational potential w calculated from $F(b,V)$ above, originates from entanglement of our branch with all else in the multiverse. This correction led us to a series of predictions in 2006 ~\cite{avatars} that test our theory such as: the existence of the giant void at 9 billion yrs away of size roughly $200 Mpc$; the alignement of lowest multipoles; power asymmetry; suppresion of dipole power; suppression of $\sigma_8$ by $30$percent; suppresion of power in the CMB $TT-$autocorrelations at low multipoles; enhancement of power at high multipoles; absence of SUSY breaking at Higgs energies; and the bulk flow of structure. The dark flow remains to be confirmed but all the other predictions are in perfect agreement with observations made since 2007 to present from WMAP, Planck, LHC and the Kashlinsky et al. team.

\subsection{Conclusions}
The theory of the origins of the universe from the quantum landscape multiverse described above, allowed us to derive the quantum selection criterion for the high energy origins of our universe from simply applying quantum mechanics on the landscape multiverse, without any ad-hoc assumptions and no use of anthropic arguments. Observationally we were able to derive predictions to test this theory. Since our predictions seem so far in perfect agreement with current observations they should be considered not only as tests of our theory for the origins of the universe but also the first tests of a rich and complex structure beyond, that of the multiverse.


\end{document}